\documentclass[prd,nofootinbib,superscriptaddress,preprint]{revtex4}
\usepackage[T1]{fontenc}
\usepackage{amsmath,amssymb}
\usepackage{epsfig}
\usepackage{graphicx,array}
\usepackage[usenames,dvipsnames]{color}
\usepackage{slashed}
\usepackage{comment}
\usepackage[colorlinks,citecolor=blue]{hyperref}
\usepackage{pdfpages}
\usepackage{color}
\usepackage{subfigure}

\begin{document}

\title{Froggatt-Nielsen like mechanism in the framework of Modular Symmetry for Neutrino Mass, Mixing and Leptogenesis} 
	\author{Gourab Pathak}
	\email{gourabpathak7@gmail.com}
	\affiliation{Department of Physics, Tezpur University, Napaam, Assam, India-784028}

\author{Mrinal Kumar Das}
\email{mkdas@tezu.ernet.in}
\affiliation{Department of Physics, Tezpur University, Napaam, Assam, India-784028}

\begin{abstract}
We study a neutrino mass model with Froggatt-Nielsen (FN) like modular symmetry. The FN mechanism requires an additional gauge symmetry, $U(1)_{FN}$, which is spontaneously broken at high energies. But in this work, we do not need an extra symmetry as modular weights play the role of the FN charges of the additional $U(1)_{FN}$ symmetry. We have constructed a neutrino mass model using FN-like modular symmetry in the $T'$ group. This model can accommodate neutrino oscillation parameters and also address other phenomena beyond the Standard Model, such as neutrinoless double beta decay and the baryon asymmetry of the universe.

\end{abstract}

\maketitle 

\section{Introduction}\label{sec:intro}
Despite the immense success of the Standard Model (SM), several unsolved problems remain. One of the most challenging problems is the flavor structure of the Standard Model fermions. The hierarchy between mixing angles and masses in both the quark and neutrino sectors can only be explained by a beyond the SM mechanism. Since in the SM, Yukawa couplings are free parameters, the hierarchy can not be explained in the framework of the SM.
\par
Recently, modular symmetry was proposed to solve the flavor puzzle of the SM\cite{Feruglio:2017spp}. In this framework, flavor structure can be achieved without requiring many flavon fields, and the flavor symmetry is completely broken by the VEV of the modulus $\tau$, which is a complex parameter. The Yukawa couplings are also not free parameters. The Yukawa couplings are written in terms of modular forms, which are functions of the \textit{modulus} $\tau$. They transform non-trivially under the finite modular group $\Gamma_{N}$, and $\Gamma_{N}$ is isomorphic to the non-abelian discrete group. In the literature, there are several modular symmetric models available: $\Gamma_{2} \cong S_{3}$\cite{Meloni:2023aru,Okada:2019xqk}, $\Gamma_{3} \cong A_{4}$\cite{Kashav:2021zir,Nomura:2019jxj,Behera:2020lpd,Altarelli:2005yx,Pathak:2025zdp,Pathak:2024sei}, $\Gamma_{4}\cong S_{4}$\cite{Kobayashi:2019xvz,Penedo:2018nmg,Zhang:2021olk,Wang:2019ovr}, $\Gamma_{5}\cong A_{5}$\cite{Novichkov:2018nkm,Ding:2019xna} and also double cover modular group $\Gamma_{N}'$ such as: $\Gamma_{2}'=\Gamma_{2}\cong S_{3}$, $\Gamma_{3}'\cong T'$\cite{Liu:2019khw,Mishra:2023cjc}, $\Gamma_{4}'\cong S_{4}'$\cite{Novichkov:2020eep,Liu:2020akv} and $\Gamma_{5}'\cong A_{5}'$\cite{Wang:2020lxk,Yao:2020zml,Behera:2021eut}. In this work, we are constructing a model based on the modular double covering group $T'$. 
\par
The Froggatt-Nielsen (FN) mechanism provides an elegant and economical framework for explaining the flavor structure of the quark and neutrino sectors using a horizontal $U(1)_{FN}$ symmetry\cite{Froggatt:1978nt,Ibe:2024cvi,Kamikado:2008jx,Qiu:2023igq,Cornella:2024jaw}. In the original FN mechanism, fermions carry additional FN charge under the gauge group $U(1)_{FN}$, and Yukawa couplings are forbidden at the tree level. The $U(1)_{FN}$ symmetry is spontaneously broken by an extra scalar field $\phi$, which is a trivial singlet under the SM gauge group. At low energies, the $U(1)_{FN}$ symmetry is spontaneously broken, and after integrating out the heavy fields, we get effective Yukawa couplings suppressed by its vacuum expectation value $(\langle\phi \rangle/\Lambda)^{n}$. Here, $\langle\phi \rangle$ is the VEV of the scalar $\phi$ and n is the difference of $U(1)_{FN}$ charges between fermion generations. This mechanism has been extensively studied in the quark sector, but its application in the leptonic sector gives rise to new challenges. In the quark sector, the masses and mixing angles all arise from Dirac-like Yukawa couplings. However, the underlying mechanism for neutrino mass generation is currently unknown, which leads to a large variety of possible implementations. In this work, we studied the FN mechanism for the leptonic sector using modular $T'$ groups, and neutrino mass is generated via the Type-I seesaw mechanism\cite{Brdar:2019iem,Behera:2024vfv,minkowski1977mu,schechter1980neutrino,mohapatra1980neutrino}. 
\par
In this work, we have used an FN-like mechanism in the framework of modular symmetry to study neutrino oscillation parameters and other beyond the SM phenomenology. The FN-like mechanism has been recently used for the quark sector\cite{King:2020qaj,Kuranaga:2021ujd}. Here we are dealing with an FN-like mechanism for the leptonic sector. In analogy to the FN mechanism, the modular weights of fermion fields serve the role of $U(1)_{FN}$ charge. The extra scalar field $\phi$, which is a singlet under the SM gauge group, compensates for the modular weights of the fermion fields. The Yukawa couplings are also of modular forms, and the effective Yukawa couplings that emerge through higher-dimensional operators are suppressed by some powers of $\Phi =(\langle \phi \rangle/\Lambda)$. $\Phi$ is called spurion associated with the breaking of $U(1)_{FN}$. 
\par
One of the outstanding problems of neutrino physics is whether neutrinos are Dirac or Majorana in nature. Neutrinoless double beta decay $(0\nu\beta\beta)$ \cite{Jones:2021cga,Dolinski:2019nrj,Bilenky:2012qi,Gomez-Cadenas:2010zcc} is a hypothetical nuclear reaction that emits two electrons without accompanying neutrinos. The observation of this reaction would confirm that neutrinos are Majorana particles, i.e., they are their own antiparticles. In this work, we have studied our model prediction for this reaction.
\par
Numerous observational findings indicate an imbalance between matter and antimatter in our universe. The SM doesn't have enough ingredients to explain this imbalance, and we need to go beyond the SM to explain this. This longstanding matter antimatter asymmetry puzzle is known as the Baryon Asymmetry of Universe (BAU). For the dynamical generation of baryon asymmetry, the three Sakharov conditions need to be satisfied\cite{Sakharov:1967dj}: \textit{Baryon number (B) violation}, \textit{C or CP violation}, and \textit{Out of equilibrium condition}. In the seesaw framework, all three conditions are satisfied naturally. Therefore, the process responsible for generating the baryon asymmetry, known as leptogenesis, becomes an integral part of the seesaw framework. In this work, we will establish the baryon asymmetry of the universe via \textit{thermal leptogenesis} in a FN-like modular model.
\par
This work is organized as follows. In section \ref{sec:2}. We briefly addressed the modular symmetry and the $T'$ group. We explicitly addressed all the components of the model framework in \ref{sec:3}. Numerical analysis for the neutrino sector, neutrinoless double beta decay, and thermal leptogenesis is discussed in section \ref{sec:4}. Finally, we concluded our analysis in section \ref{sec:5}.

\section{MODULAR SYMMETRY AND $T'$ GROUP}\label{sec:2}
\subsection{MODULAR SYMMETRY}

$\bar{\Gamma}$ is the modular group that attains a linear fractional transformation $\gamma$ which acts on modulus $\tau$ linked to the upper-half complex plane whose transformation is given by\cite{Feruglio:2017spp}:
\begin{equation}
\gamma \longrightarrow \gamma \tau = \frac{a\tau + b}{c\tau + d},
\end{equation}
where $a, b, c, d \in  \mathbb{Z}$ and $ad-bc = 1$, $Im[\tau]>0$, where it is isomorphic to the transformation $PSL(2, \mathbb{Z}) = SL(2, \mathbb{Z})/\{I, -I\}$. The S and T transformation helps in generating the modular transformation defined by:
\begin{equation}
    S: \tau \longrightarrow -\frac{1}{\tau}, \hspace{1.5cm} T: \tau \longrightarrow \tau + 1,
\end{equation}
and hence the algebraic relations so satisfied are as follows,
\begin{equation}
    S^2 = \mathbb{I}, \hspace{2.5cm} (ST)^3 = \mathbb{I}.
\end{equation}
Here, series of groups are introduced, $\Gamma(N) (N = 1, 2, 3, .....)$ and defined as
\begin{equation}
    \Gamma(N) =\Bigg \{\begin{pmatrix}
        a & b \\
        c & d
    \end{pmatrix} \in SL(2,\mathbb{Z}), \begin{pmatrix}
        a & b \\
        c & d 
    \end{pmatrix} = \begin{pmatrix}
        1 & 0 \\
        0 & 1
    \end{pmatrix}(\text{mod} N)\Bigg\}
\end{equation}
Definition of $\bar{\Gamma}(2) \equiv \Gamma(2)/\{I, -I\}$ for N=2. Since $-I$ is not associated with $\Gamma(N)$ for $N>2$ case, one can have $\bar{\Gamma}(N)=\Gamma(N)$, which are infinite normal subgroups of $\bar{\Gamma}$ known as principal congruence subgroups. Quotient groups come from the finite modular group, defined as $\Gamma_N = \bar{\Gamma}/\bar{\Gamma}(N)$. The
imposition of $T^N = \mathbb{I}$ is done for these finite groups $\Gamma_N$. Thus, the groups $\Gamma_N (N = 2, 3, 4, 5)$ are isomorphic to $S_3$, $A_4$, $S_4$ and $A_5$, respectively. N level modular forms are holomorphic functions $f(\tau)$ which are transformed under the influence of $\Gamma(N)$ as follows:
\begin{equation}
    f(\gamma\tau)=(c\tau + d)^{k}f(\tau), \hspace{1.5cm} \gamma \in \Gamma(N)
\end{equation}
where k is the modular weight. The prefactor $(c\tau+d)^k$ is referred to as the automorphy factor. Since linear combinations of the modular forms of level N and weight k are also modular forms of level N and weight k. Hence they form a linear space $\mathcal{M}_{k}(\Gamma_{N})$. The  modular forms from the unitary representations of the quotient group $\Gamma_{N}=\Gamma/\Gamma(N)$ are given by: 
\begin{equation}
    f_{i}(\gamma\tau)=(c\tau + d)^{k}\rho(\gamma)_{ij}f_{j}(\tau)
\end{equation}is 
where ${f_{i}}$ is a basis of $\mathcal{M}_{k}(\Gamma_{N})$, and $\rho$ is a unitary representation of $\Gamma_{N}$. From the automorphy factor, it can be seen that the modular group is divided by its center, and therefore, modular forms are well defined only for even modular weight. The $\Gamma$ is called the inhomogeneous modular group to distinguish it from the double covering group of the modular group $\Gamma'$, which is homogeneous. Similar to $\Gamma_N$ its quotient group $\Gamma_{N}'$ is obtained as follows:
\begin{equation}
    \Gamma_{N}'=\Gamma'/\Gamma'(N),
\end{equation}
where $\Gamma'(N)$ is a subgroup of $\Gamma'$. $\Gamma'$ is nothing but $SL(2, \mathbb{Z})$ itself. The following relations are satisfied by the generators of $\Gamma_{N}'$:
\begin{equation}
    S^2=R,\quad (ST)^{3}=I,\quad T^{N}=I,\quad TR=RT,\quad R^2=I.
\end{equation}
Similar to $\Gamma_{N}$, for $N < 6$ the $\Gamma_{N}'$ is also isomorphic to non-Abelian group: $\Gamma_{2}'\simeq S_3$, $\Gamma_{3}'\simeq T'$, $\Gamma_{4}'\simeq S_{4}'$ and $\Gamma_{5}'\simeq A_{5}'$. In the double covering of the modular group, we can define modular forms of odd modular weights as well.  
In this work, we consider a modular symmetric model based on $\Gamma'$, and all the matter fields are represented by chiral superfields. These fields transform as the modular forms of level N and weight k as follows:
\begin{equation}
    \gamma\quad: \quad \Phi_{i} \rightarrow (c\tau + d)^{k_{i}}\rho_{\Phi}(\gamma)_{i,j}\Phi_{j}, 
\end{equation}
where $\Phi_{i}$ is a matter field and $\rho_{\Phi}(\gamma)$ is a unitary matrix. The action of chiral superfields $\Phi_{i}$ is given by:
\begin{equation}
    \mathcal{S}=\int{d^{4}xd^{2}\theta d^{2}\bar{\theta}K(\Phi^{i}, \bar{\Phi}^{i}, \tau, \bar{\tau})} + \int{d^{4}xd^{2}\theta W(\Phi^{i}, \tau)} + h.c,
\end{equation}
where $K$ and $W$ are the K$\ddot{a}$hler potential and the superpotential respectively. The modular invariant $K$ and $W$ is given as follows:
\begin{equation}
\begin{split}
    K=&\sum_{i}\frac{\Phi_{i}\bar{\Phi}_{i}}{Im\tau^{-k_{i}}}\\
    W=&\sum(f_{i_{1}i_{2}.....i_{n}}(\tau)\Phi_{i_{1}}\Phi_{i_{2}}.....\Phi_{i_n})_{1}
\end{split}
\end{equation}
where  $f_{i_{1}i_{2}.....i_{n}}(\tau)$ is a modular form of weight $k$ satisfying:
\begin{equation}
    k+k_{i_{1}}+k_{i_{2}}+.....+k_{i_{n}}=0,
\end{equation}
so that the modular weight of the superpotential is zero. The allowed Yukawa couplings are also given in the modular forms so that they remain invariant under modular symmetry.
\subsection{$T'$ GROUP}
In the section above, we have discussed that $\Gamma_{3}'$ is isomorphic to $T'$\cite{Ding:2022aoe,Liu:2019khw,Kuranaga:2021ujd}. This group is a double covering of the tetrahedral group $A_4$. The $T'$ has 24 elements, and all these elements can be generated by the three generators S, T, and R  fulfilling the following relations:
\begin{equation}
    S^2=R,\quad (ST)^{3}=I, \quad T^{3}=I, \quad R^2=I,\quad RT=TR.
\end{equation}
This group has seven irreducible representations, which are mentioned below:
\begin{equation}
    1,\quad 1',\quad 1'',\quad 2,\quad 2',\quad 2'',\quad 3.
\end{equation}
The multiplication rule for this representation under $T'$ is:
\begin{equation}
    \begin{split}
        1 \otimes 1 &= 1' \otimes 1'' = 1,\quad 1 \otimes 1' = 1'' \otimes 1'' = 1',\quad 1 \otimes 1'' = 1' \otimes 1' = 1'',\\
        2 \otimes 2 &= 2'\otimes2''=3\oplus1',\quad 2\otimes2'=2''\otimes2''=3\oplus1'',\quad 2\otimes2''=2'\otimes2'=3\oplus1\\
        2\otimes3&=2\oplus2'\oplus2'',\quad
        2'\otimes3=2\oplus2'\oplus2'',\quad 
        2\otimes3=2\oplus2'\oplus2''\\
        3\otimes3 &= 1 \oplus 1' \oplus 1'' \oplus 3_S \oplus 3_A.
    \end{split}
\end{equation}

For the modular form of weight 1 and level 3, we can construct a doublet of $\Gamma_{3}' \cong T'$:
\begin{equation}
    Y_{2}^{(1)} = \begin{pmatrix}
        Y_{1}(\tau)\\
        Y_{2}(\tau)
    \end{pmatrix}
\end{equation}

The q-expansion of the component of the doublet modular form $Y_{2}^{(1)}$ is given by:
\begin{equation}
\begin{split}
    Y_{1}(\tau) &= \sqrt{2}e^{i7\pi/12}q^{1/3}(1+q+2q^2+2q^4+q^5+2q^6+.......),\\
    Y_{2}(\tau) &= 1/3 + 2q + 2q^3 + 2q^4 + 4q^7 +2q^9+.........
\end{split}    
\end{equation}
where $q=e^{i2\pi\tau}$. The detailed derivation of this can be found in \cite{Kuranaga:2021ujd}. Higher modular weight Yukawa couplings can be constructed from the weight 1 Yukawa coupling $Y_2^{(1)}$ using the $T'$ multiplication rule. For  modular weight $k=2$, we have the following modular forms:
\begin{equation}
\begin{split}
    Y_{1}^{(2)}&=\bigg(Y_{2}^{(1)}Y_{2}^{(1)}\bigg)_{1'}=Y_{1}Y_{2}-Y_{2}Y_{1}=0,\\
    Y_{3}^{(2)}&=(Y_{2}^1Y_{2}^1)_{3}=\bigg(e^{i\pi/6}Y_{2}^{2},\quad \sqrt{2}e^{i7\pi/12}Y_{1}Y_{2},\quad Y_{1}^{2}\bigg)^{T}
    \end{split}
\end{equation}
For modular weight 3, we can have the following modular forms:
\begin{equation}
    \begin{split}
        Y_{2}^{3}&=\bigg(Y_{2}^{(1)}Y_{3}^{(2)}\bigg)_{2}=(3e^{i\pi/6} Y_{1}Y_{2}^{2},\quad\sqrt{2}e^{i5\pi/12}Y_{1}^{3}-e^{i\pi/6}Y_{2}^{3})^{T},\\
        Y_{2'}^{(3)}&=\bigg(Y_{2}^{(1)}Y_{3}^{(2)}\bigg)_{2'}=(0,\quad0)^{T},\\
        Y_{2''}^{(3)}&=\bigg(Y_{2}^{(1)}Y_{3}^{(2)}\bigg)_{2''}=(Y_{1}^{3}+(1-i)Y_{2}^{3},\quad -3Y_{2}Y_{1}^{2})^{T}.
    \end{split}
\end{equation}
Similarly, for weight 4, the following non-vanishing and independent modular forms can be constructed:
\begin{equation}
    \begin{split}
        Y_{3}^{(4)}&=\bigg(Y_{2}^{(1)}Y_{2}^{(3)}\bigg)_{3}=(\sqrt{2}e^{i7\pi/12}Y_{1}^{3}Y_{2}-e^{i\pi/3}Y_{2}^{4},\quad -Y_{1}^{4}-(1-i)Y_{1}Y_{2}^{3},\quad 3e^{i\pi/6}Y_{1}^{2}Y_{2}^{2})^{T},\\
        Y_{1'}^{(4)}&=\bigg(Y_{2}^{(1)}Y_{2}^{(3)}\bigg)_{1'}=\sqrt{2}e^{i5\pi/12}Y_{1}^{4}-4e^{i\pi/6}Y_{1}Y_{2}^{3},\\
        Y_{1}^{(4)}&=-4Y_{1}^{3}Y_{2}-(1-i)Y_{2}^{4}.
         \end{split}
\end{equation}
The following relations give the weight 5 non-vanishing and independent modular forms:
\begin{equation}
    \begin{split}
        Y_{2}^{(5)}&=\bigg(Y_{2}^{(1)}Y_{3}^{(4)}\bigg)_{2}=(2\sqrt{2}e^{i7\pi/12}Y_{1}^{4}Y_{2}+e^{i\pi/3}Y_{1}Y_{2}^{4},\quad 2\sqrt{2}e^{i7\pi/12}Y_{1}^{3}Y_{2}^{2}+e^{i\pi/3}Y_{2}^{5})^{T},\\
        Y_{2'}^{(5)}&=\bigg(Y_{2}^{(1)}Y_{3}^{(4)}\bigg)_{2'}=(-Y_{1}^{5}+2(1-i)Y_{1}^{2}Y_{2}^{3},\quad -Y_{1}^{4}Y_{2}+2(1-i)Y_{1}Y_{2}^{4})^{T},\\
        Y_{2''}^{(5)}&=\bigg(Y_{2}^{(1)}Y_{3}^{(4)}\bigg)_{2''}=(5e^{i\pi/6}Y_{1}^{3}Y_{2}^{2}-(1-i)e^{i\pi/6}Y_{2}^{5},\quad -\sqrt{2}e^{i5\pi/12}Y_{1}^{5}-5e^{i\pi/6}Y_{1}^{2}Y_{2}^{3})^{T}.
    \end{split}
\end{equation}
We can construct other higher weight modular forms similar way. But in this work, we are dealing with up to weight 5 modular forms.

\section{MODEL FRAMEWORK}\label{sec:3}

\begin{table}[h]
    \centering
    \begin{tabular}{||c|c|c|c|c|c|c|c|c|c||}
    \hline
    Symmetry & \multicolumn{9}{c||}{Superfield content and charge} \\
    \hline
    & $L$ & $e_R^c$ & $\mu_R^c$ & $\tau_R^c$ & $N_{R_1,R_2}^c$ & $N_{R_3}^c$ & $H_{u}$ & $H_{d}$  & $\phi$ \\
    \hline
    $SU(2)_L$ & 2 & 1 & 1 & 1 & 1 & 1 & 2 & 2 & 1  \\
    $U(1)_Y$ & -1 & 2 & 2 & 2 & 0 & 0 & 1 & -1 & 0 \\
    $T'$ & $3$ & $1$ & $1''$ & $1'$ & $2$ & 1 & $1$ & $1$ & 1  \\
    $k_{I}$ & -2 & 3 & 3 & 3 & 2 & 4 & 0 & 0 & 0  \\
    \hline
    \end{tabular}
    \caption{Superfield content of the model with their $T'$ and modular charges $k_I$.}
    \label{tab:particle-content}
\end{table}

\begin{table}[h]
    \centering
    \begin{tabular}{||c|c|c|c|c||}
    \hline
    & \multicolumn{4}{c||}{Yukawa Couplings} \\ \hline
    & $Y_{3}^{(2)}$ & $Y_{2}^{(3)}$ & $Y_{2''}^{(5)}$ & $Y_{2''}^{(3)}$  \\ \hline
    $T'$ & $3$ & $2$ & $2''$ & $2''$ \\ 
    $k_{I}$ & $2$ & $3$ & $5$ & $3$  \\ \hline
    \end{tabular}
    \caption{Yukawa couplings with their $T'$ and modular charges $k_I$.}
    \label{tab:yukawa-content}
\end{table}

    In this work, we have constructed a model based on modular $\Gamma_{3}^{'} \equiv T'$. To study the Type I seesaw mechanism for neutrino mass generation, we have considered the charged assignment for different superfields as given in the Table \ref{tab:particle-content}. The different modular weight Yukawa couplings used for this analysis is given in the Table \ref{tab:yukawa-content}. We assign $T'$ triplet \textbf{3} to the three generation of left-handed lepton doublet $L$ while the three right-handed charged lepton $e_{R}^c$, $\mu_{R}^{c}$ and $\tau_{R}^{c}$ transform as \textbf{1}, $\textbf{1}''$ and $\textbf{1}'$ respectively under $T'$. Two of the three right-handed neutrinos $(N_{R_1}^{c}, N_{R_2}^{c})$ are embedded into a $T'$ doublet \textbf{2} while the remaining $N_{R_3}^{c}$ is assigned as \textbf{1}. We also assign trivial singlet \textbf{1} under $T'$ to both Higgs doublets $H_{u/d}$. We also introduce a new chiral superfield $\phi$ called weighton. This $\phi$ is the trivial singlet under both $T'$ and the SM gauge group. It also carries -1 modular weight. The modular weight for other fields is presented in the Table \ref{tab:particle-content}. After the breaking of modular symmetry, $\phi$ develops its VEV, and the effective superpotentials for our FN-like model with $T'$ obtained are discussed below. In all the superpotential terms, $\Phi$ stands for $\langle \phi \rangle/\Lambda$, where $\Lambda$ is the cut-off scale. In this study, we have taken the FN scale the same as the Majorana mass scale $M=\Lambda$ \cite{Cornella:2024jaw}.

\par

The charge assignment is shown in Table \ref{tab:particle-content} gives the following charged lepton superpotential:
\begin{equation}
    \mathcal{W}_l = \alpha e_{R}^cH_{d}\bigg(LY_{3}^{(2)}\bigg)_{1}\Phi^3 + \beta \mu_{R}^cH_{d}\bigg(LY_{3}^{(2)}\bigg)_{1'}\Phi^3 + \gamma\tau_{R}^cH_{d}\bigg(LY_{3}^{(2)}\bigg)_{1''}\Phi^3
\end{equation}
The charged lepton mass matrix takes the form:
\begin{equation}
    M_{l} = v_d\Phi^3
    \begin{pmatrix}
        \alpha e^{\frac{i\pi}{6}}Y_{2}Y_{2} & \alpha Y_1Y_1 & \alpha\sqrt{2}e^{\frac{i7\pi}{12}Y_1Y_2}\\
        \beta\sqrt{2}e^{\frac{i7\pi}{12}}Y_1Y_2 & \beta e^{\frac{i\pi}{6}}Y_2Y_2 & \beta Y_1Y_1\\
        \gamma Y_1Y_1 & \gamma \sqrt{2}e^{\frac{i7\pi}{12}}Y_1Y_2 & \gamma e^{\frac{i\pi}{6}Y_2Y_2}
    \end{pmatrix}
    \label{Eq: charged lepton}
\end{equation}    

Here, $\langle H_{d} \rangle = v_d$ is the VEV of Higgs superfield $H_{d}$ and $\{\alpha, \beta, \gamma\}$ are free parameters. The charged lepton mass matrix is found to be non-diagonal.

\par

The invariant Dirac superpotential involving lepton fields $L_i$  and heavy right-handed fields $N_{R_{i}}^c$ is given as follows: 
\begin{equation}
    \mathcal{W}_{D} = g_{1}\bigg(\bigg(N_{R_1,R_2}^cL\bigg)_{2''}Y_{2}^{(3)}\bigg)_{1}H_{u} + g_2\bigg(\bigg(N_{R_1,R_2}^cL\bigg)_{2}Y_{2''}^{(3)}\bigg)_{1}H_{u} + g_{3}\bigg(\bigg(N_{R_3}^{c}L\bigg)_{3}Y_{3}^{(3)}\bigg)_{1}H_{u}
\end{equation}
The Dirac mass matrix is obtained to be:
\begin{equation}
    M_{D} = v_u\begin{pmatrix}
        m_{11} & m_{12} & m_{13}\\
        m_{21} & m_{22} & m_{23} \\
        m_{31} & m_{32} & m_{33}
    \end{pmatrix}
\end{equation}
where the elements of the matrix are given as follows:
\begin{equation}
\begin{split}
    m_{11}& = g_{2}\Phi^{5}\bigg(5e^{\frac{i\pi}{6}}(Y_{1})^{3}(Y_{2})^{2}-(1-i)e^{\frac{i\pi}{6}}(Y_2)^{5}\bigg)\\
    m_{12}& = \bigg[g_{1}\Phi^{3}\bigg(-3\sqrt{2}e^{\frac{i9\pi}{12}}Y_{1}(Y_2)^{2}\bigg)+g_{2}\Phi^{5}\bigg(\sqrt{2}e^{\frac{i5\pi}{12}}(Y_1)^{5}+5e^{\frac{i\pi}{6}}(Y_1)^{2}(Y_2)^{3}\bigg)\bigg]\\
    m_{13}& = g_{3}\Phi^{4}e^{\frac{i\pi}{6}}(Y_2)^{2}\\
    m_{21}& = g_{1}\Phi^{3}\bigg(2e^{\frac{i5\pi}{6}}(Y_1)^{3}-\sqrt{2}e^{\frac{i7\pi}{12}}(Y_2)^{3}\bigg)\\
    m_{22}& = g_{2}\Phi^{5}\bigg[(1-i)\sqrt{2}e^{\frac{i9\pi}{12}}(Y_2)^{5}-5\sqrt{2}e^{\frac{i9\pi}{12}}(Y_1)^{3}(Y_2)^{2}\bigg]\\
    m_{23}& = g_{3}\Phi^{4}\sqrt{2}e^{\frac{i7\pi}{12}}Y_1Y_2\\
    m_{31}& = \bigg[g_{1}\Phi^{3}\bigg(3e^{\frac{i\pi}{6}}Y_{1}(Y_2)^{2}\bigg)-g_{2}\Phi^{5}\bigg(2e^{\frac{i5\pi}{6}}(Y_1)^{5}+5\sqrt{2}e^{\frac{i7\pi}{12}}(Y_1)^{2}(Y_2)^{3}\bigg)\bigg]\\
    m_{32}& = g_{1}\Phi^{3}\bigg(e^{\frac{i\pi}{6}}(Y_2)^{3}-\sqrt{2}e^{\frac{i5\pi}{12}}(Y_1)^{3}\bigg)\\
    m_{33}& = g_{3}\Phi^{4}(Y_1)^{2}
\end{split}    
\end{equation}

where $\{g_1,g_2,g_3\}$ are free parameters and $\langle H_u \rangle = v_u$ is the VEV of Higgs superfield $H_u$.

\par

The superpotential involving the Majorana mass term can be written as follows:
\begin{equation}
    \mathcal{W}_{R} = \Lambda\bigg(N_{R_1,R_2}^{c}N_{R_1,R_2}^{c}\bigg)_{3}Y_{3}^{(2)}\Phi^6+\Lambda\bigg(N_{R_3}^{c}N_{R_1.R_2}^{c}\bigg)_{2}Y_{2''}^{(3)}\Phi^{9}+\Lambda\bigg(N_{R_3}^{c}N_{R_3}^{c}\bigg)\Phi^8
\end{equation}
where $\Lambda$ is the mass scale.\\
This gives the following Majorana mass matrix:
\begin{equation}
    M_{R} = \Lambda\begin{pmatrix}
        \sqrt{2}e^{\frac{i7\pi}{12}}Y_1Y_2\Phi^{6} & \frac{1}{\sqrt{2}}e^{\frac{i7\pi}{12}}(Y_1)^{2}\Phi^{6} & -3(Y_1)^{2}Y_2\Phi^{9}\\
        \frac{1}{\sqrt{2}}e^{\frac{i7\pi}{12}}(Y_1)^{2}\Phi^{6} & e^{\frac{i\pi}{3}(Y_2)^{2}\Phi^{6}} & -((Y_1)^{3}+(1-i)(Y_2)^{3})\Phi^{9}\\
        -3(Y_1)^{2}Y_2\Phi^{9} & -((Y_1)^{3}+(1-i)(Y_2)^{3})\Phi^{9} & \Phi^{8}
        
    \end{pmatrix}
\end{equation}

Within the current model, which incorporates \(T'\) modular symmetry, the complete \(6 \times 6\) neutral fermion mass matrix for the Type I seesaw mechanism in the flavor basis \((\nu_{L}, N_{R}^c)\) is given by

\begin{eqnarray}
   \mathbb{M}=  \begin{pmatrix}
        0&M_D\\M_D^T&M_R\\
    \end{pmatrix}.\label{eq:M matrix}
\end{eqnarray}

Using the appropriate mass hierarchy among the mass matrices as provided below:
\begin{equation}
    M_{R}>>M_{D},
\end{equation}
The light neutrino mass in the Type I seesaw is given by:
\begin{equation}
    m_{\nu} = M_D{M_R}^{-1}M_D^T.
    \label{eq:nu mass}
\end{equation}

\section{Numerical analysis}\label{sec:4}
\subsection{Neutrino fit data}

In general, the parameters $\{\alpha, \beta, \gamma\}$ taken in the charged lepton mass matrix in Eq.\eqref{Eq: charged lepton} are complex. However, they can be taken as real without the loss of generality by absorbing their phases into the right-handed lepton fields. Hence, their value can be completely determined by the charged lepton masses via the following relations:
\begin{equation}
    \begin{split}
        Tr(M_{l}M_{l}^{\dagger})& = m_{e}^2 +m_{\mu}^2 +m_{\tau}^2\\
        Det(M_{l}M_{l}^{\dagger})& = m_{e}^{2}m_{\mu}^{2}m_{\tau}^{2}\\
        \frac{1}{2}[Tr(M_{l}M_{l}^{\dagger})]^{2}-\frac{1}{2}Tr[(M_{l}M_{l}^{\dagger})^{2}]& = m_{e}^{2}m_{\mu}^{2}+m_{\mu}^{2}m_{\tau}^{2}+m_{\tau}^{2}m_{e}^{2}
    \end{split}
\end{equation}
Also, the charged lepton mass matrix is non-diagonal in this study, and it is diagonalized by the unitary matrix $U_{l}$, with $U_{l}^{\dagger}M_{l}M_{l}^{\dagger}U=diag(m_{e}^2,m_{\mu}^2,m_{\tau}^2)$.
The neutrino mass matrix obtain in Eq.\eqref{eq:nu mass} is diagonalized by using the relation: $U_{\nu}^{\dagger}\mathcal{M}U_{\nu} = diag(m_{1}^{2}, m_{2}^{2}, m_{3}^{2})$, where $\mathcal{M} = m_{\nu}m_{\nu}^{\dagger}$ and $U_{\nu}$ is a unitary matrix. Hence, the well-known PMNS matrix is calculated using the relation $U=U_{l}^{\dagger}U_{\nu}$ \cite{hochmuth2007upmns}. This matrix is parametrized by three mixing angles $(\theta_{12}, \theta_{13}, \theta_{23})$, one Dirac phase ($\delta_{CP}$) and two Majorana phases ($\alpha_{21}, \alpha_{31}$) as given below:

\begin{equation}
 U_{PMNS} = \begin{pmatrix}
     c_{12}c_{13} & s_{12}c_{13} & s_{13}e^{-i\delta_{CP}}\\
     - s_{12}c_{23} - c_{12}s_{23}s_{13}e^{i\delta_{CP}} & c_{12}c_{23} - s_{12}s_{23}s_{13}e^{i\delta_{CP}} & s_{23}c_{13}\\
     s_{12}s_{23} - c_{12}c_{23}s_{13}e^{i\delta_{CP}} & - c_{12}s_{23} - s_{12}c_{23}s_{13}e^{i\delta_{CP}} & c_{23}c_{13}
 \end{pmatrix}  
 . \begin{pmatrix}
     1 & 0 & 0\\
     0 & e^{i\frac{\alpha_{21}}{2}} & 0\\
     0 & 0 & e^{i\frac{\alpha_{31}}{2}}
 \end{pmatrix}
\end{equation}
where, \(c_{ij}=\cos{\theta_{ij}}\) and \(s_{ij}=\sin{\theta_{ij}}\). Using the PMNS matrix, we can calculate the three neutrino mixing angles as follows:
\begin{equation}
    \sin^2{\theta_{13}}=|U_{13}|^2, \, \sin^2{\theta_{12}} = \frac{|U_{12}|^2}{1 - |U_{13}|^2}, \, \sin^2{\theta_{23}}=\frac{|U_{23}|^2}{1 - |U_{13}|^2}.
\end{equation}
The Jarlskog invariant $J_{CP}$ and the CP-violating phase $\delta_{CP}$ are calculated using the elements of the PMNS matrix as given below:
\begin{equation}
    J_{CP} = Im[U_{e1}U_{\mu2}U_{e2}^{*}U_{\mu1}^{*}] = s_{23}c_{23}s_{12}c_{12}s_{13}c_{13}^{2}\sin{\delta_{CP}}.
\end{equation}

In numerical analysis, we use the global fit neutrino oscillation data within a $3\sigma$ interval as given in Table \ref{tab:oscillation data}.
\begin{table}[h]
    \centering
    \begin{tabular}{||c||c|c||}
    \hline
    \multicolumn{3}{|c||}{Normal ordering}\\
    \hline
           & bfp $\pm 1\sigma$ & $ 3\sigma$ range \\
         \hline
         $\sin^2{\theta_{12}}$ & $0.303_{-0.011}^{+0.012}$  & $0.270 - 0.341$ \\
         $\sin^2{\theta_{13}}$& $0.02203_{-0.00059}^{+0.00056}$ & $0.02029 - 0.02391$ \\
         $\sin^2{\theta_{23}}$& $0.572_{-0.023}^{+0.018}$ & $0.406 - 0.620$  \\
         $\frac{\Delta m_{21}^2}{10^{-5} (eV^2)}$ & $7.41_{-0.20}^{+0.21}$ & $6.82 - 8.03$  \\
         $\frac{\Delta m_{31}^2}{10^-{3} (eV^2)}$& $2.511_{-0.027}^{+0.028}$ & $2.428 - 2.597$  \\
    \hline
    \end{tabular}
    \caption{The NuFIT 5.2 (2022) results \cite{Esteban:2020cvm}.}
    \label{tab:oscillation data}
\end{table}

\begin{figure}[!ht]
        \centering
        \includegraphics[scale=0.45]{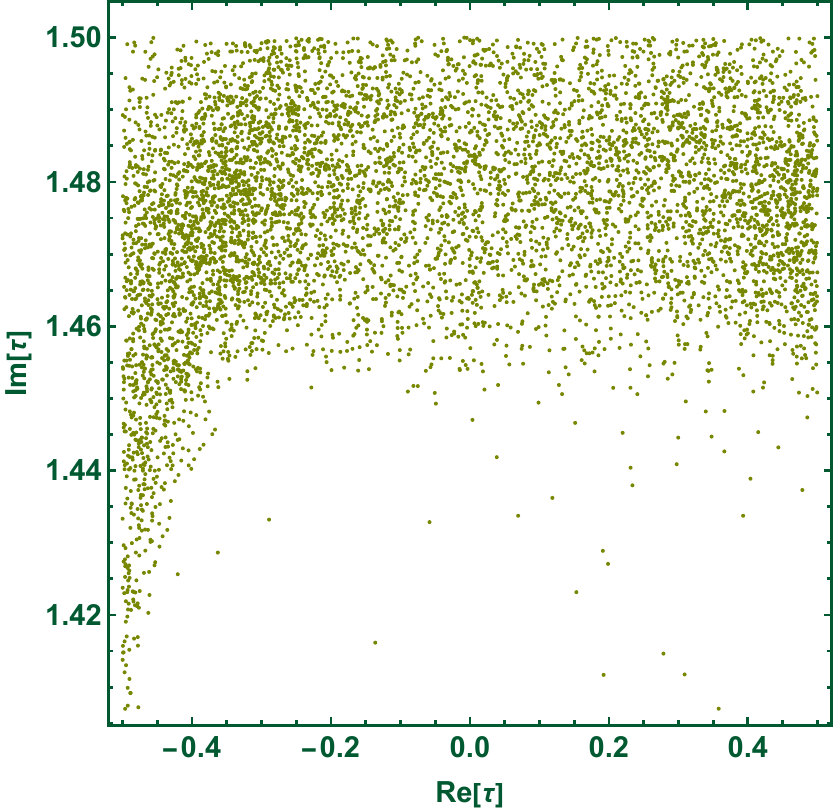}
        \caption{ Correlation between the real component and the imaginary component of $\tau$.}
       \label{fig:ReIm Tau} 
\end{figure}
\begin{figure}
    \centering
    \includegraphics[scale=0.45]{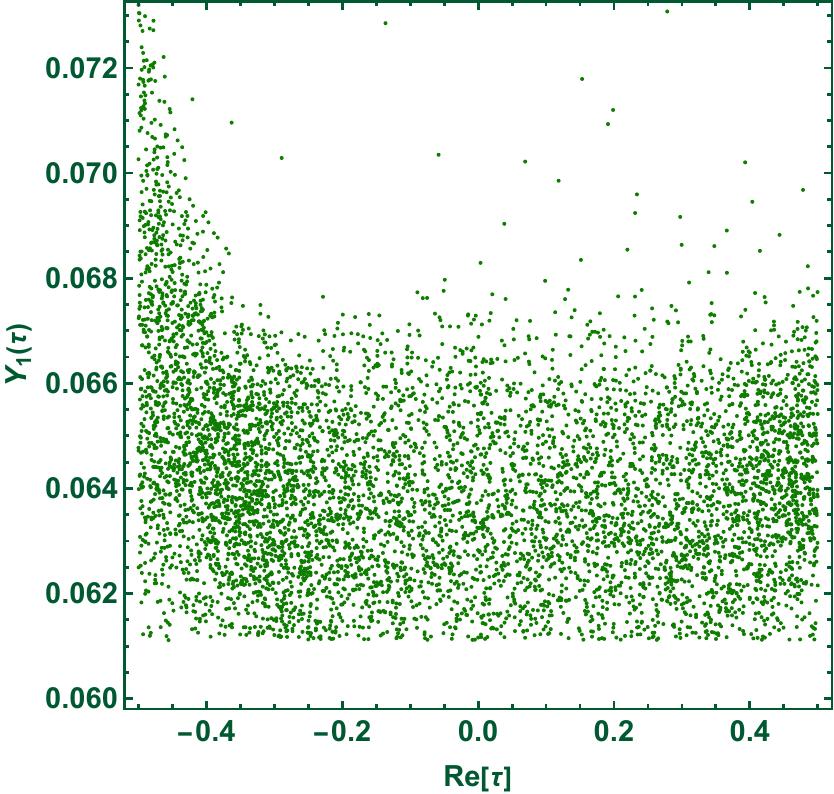}
    \hspace{2em}
    \includegraphics[scale=0.45]{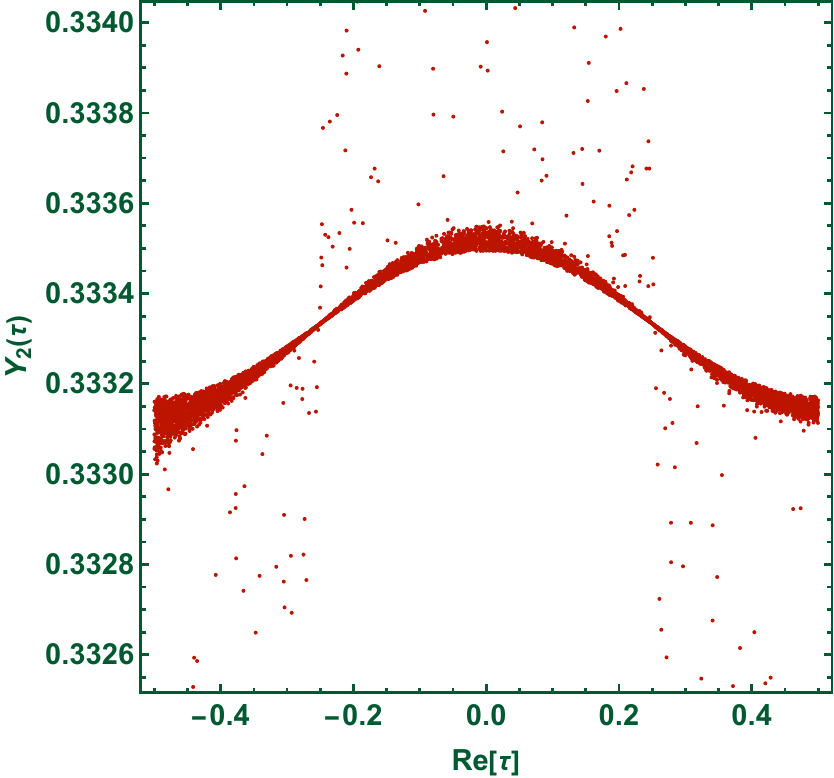}
    \includegraphics[scale=0.45]{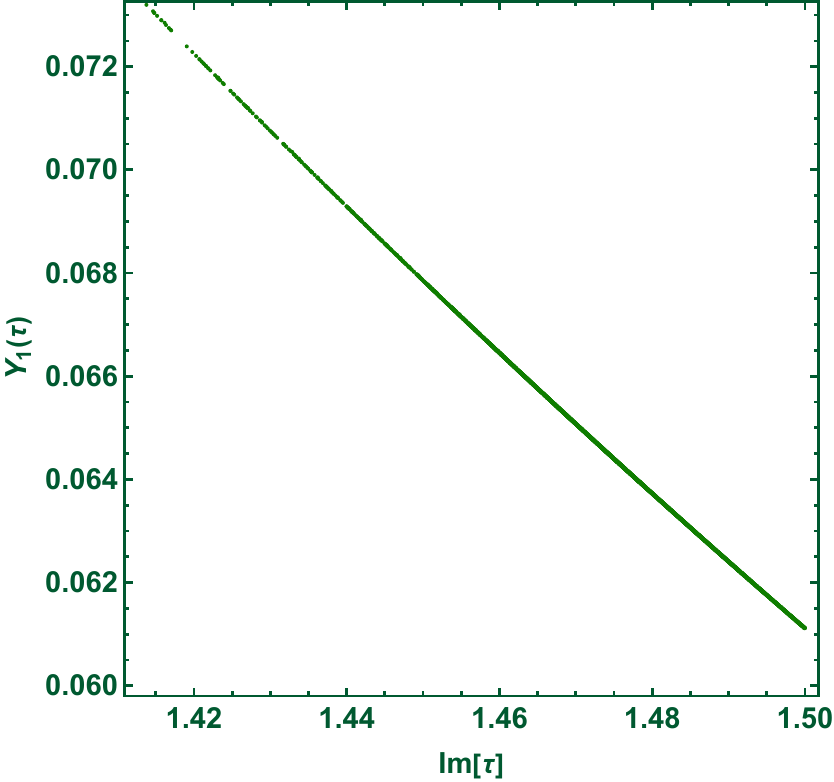}
    \hspace{2em}
    \includegraphics[scale=0.45]{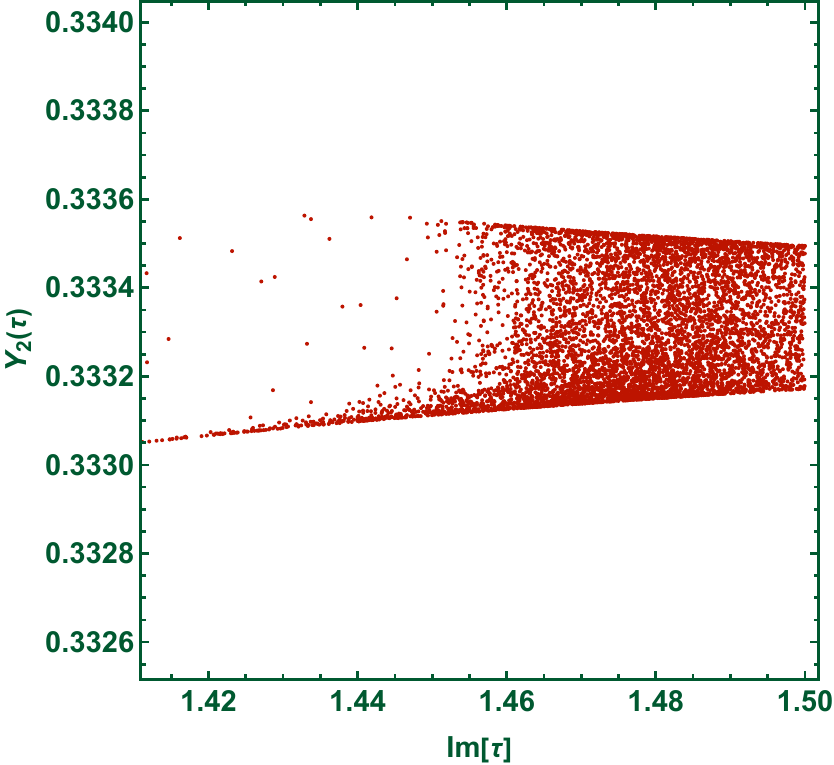}
    \caption{$(Top-panel)$ Correlation of Yukawa couplings $Y_{1}(\tau)$ and $Y_{2}(\tau)$ with the real part of modulus i.e. $Re[\tau]$. $(Bottom-panel)$ Relationship of Yukawa couplings $Y_{1}(\tau)$ and $Y_2(\tau)$ with the imaginary part of modulus i.e. $Im[\tau]$.}
    \label{fig:Yukawa vs tau}
\end{figure}
We have taken the following ranges for the model parameters to match the neutrino oscillation data:
\begin{equation}
\begin{split}
& \text{Re}[\tau] \in [-0.5, 0.5], \quad \text{Im}[\tau] \in [0.8,1.5], \quad \{g_{1}, g_{2}, g_{3} \} \in [0.1, 1],\\
& \Lambda \in [10^{11}, 10^{14}]\, GeV, \quad \Phi \in [0.2, 0.6].
\end{split}
\end{equation}
The vacuum expectation value (VEV) of the Higgs superfields, \( \langle H_u \rangle = \frac{v_u}{\sqrt{2}} \) and \(\langle H_d \rangle = \frac{v_d}{\sqrt{2}} \) are related to the Standard Model Higgs VEV, \( v_H \), through the relation \( v_H = \sqrt{v_u^2 + v_d^2} \) and the ratio of their VEVs is expressed as \( \tan\beta = \frac{v_u}{v_d} = 5 \) \cite{Kashav:2021zir,Antusch:2013jca}. 
The input parameters are randomly scanned within the specified parameter ranges. The allowed regions are selected by applying the $3\sigma$ constraints on the solar and atmospheric mass-squared differences and mixing angles as listed in the
 Table.\ref{tab:oscillation data}. The relation between the real part of modulus \(\text{Re}[\tau]\) and the imaginary part of modulus \(\text{Im}[\tau]\) is shown in Fig.\ref{fig:ReIm Tau}. The $Top-panel$ of Fig.\ref{fig:Yukawa vs tau} illustrates the correlation between the Yukawa coupling components $Y_1$, $Y_2$, and the real part of the modulus $\text{Re}[\tau]$, whereas the $Bottom-panel$ presents their correlation with the imaginary part $\text{Im}[\tau]$.

\begin{figure}
    \centering
    \includegraphics[scale=0.45]{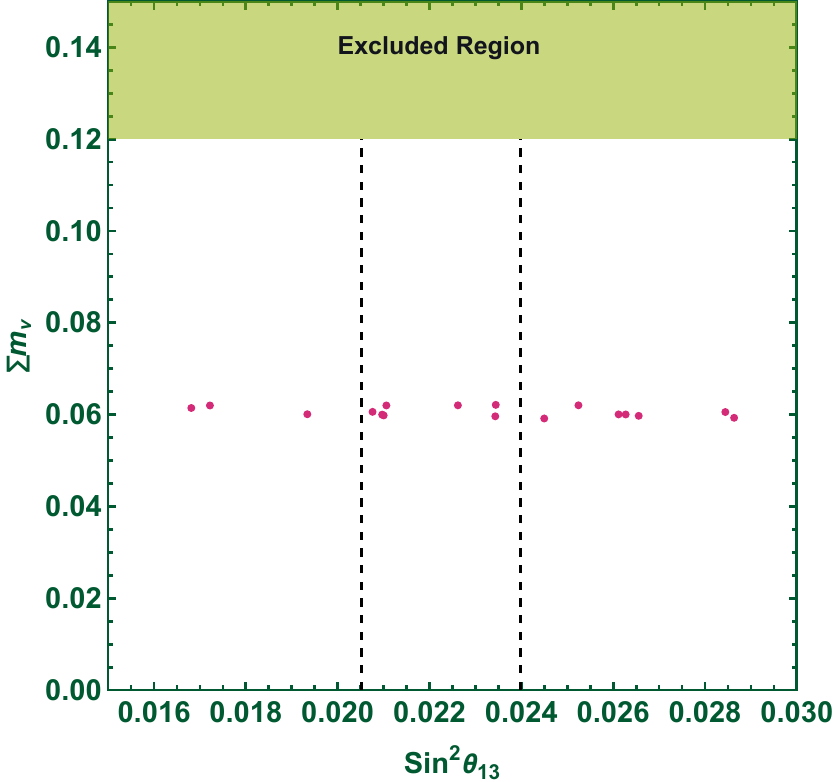}
    \hspace{2em}
    \includegraphics[scale=0.45]{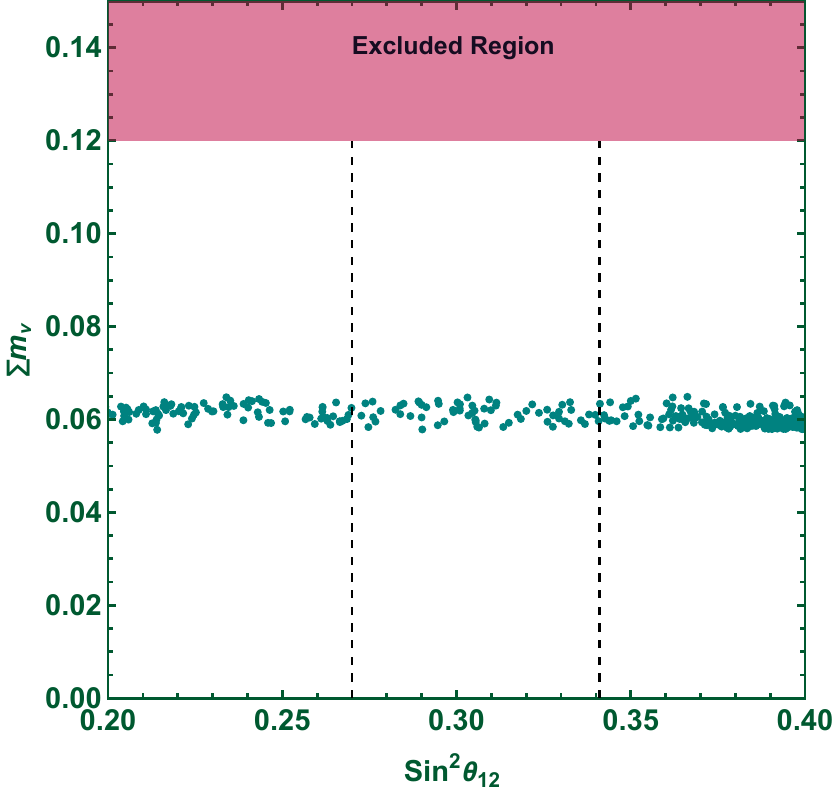}
    \includegraphics[scale=0.45]{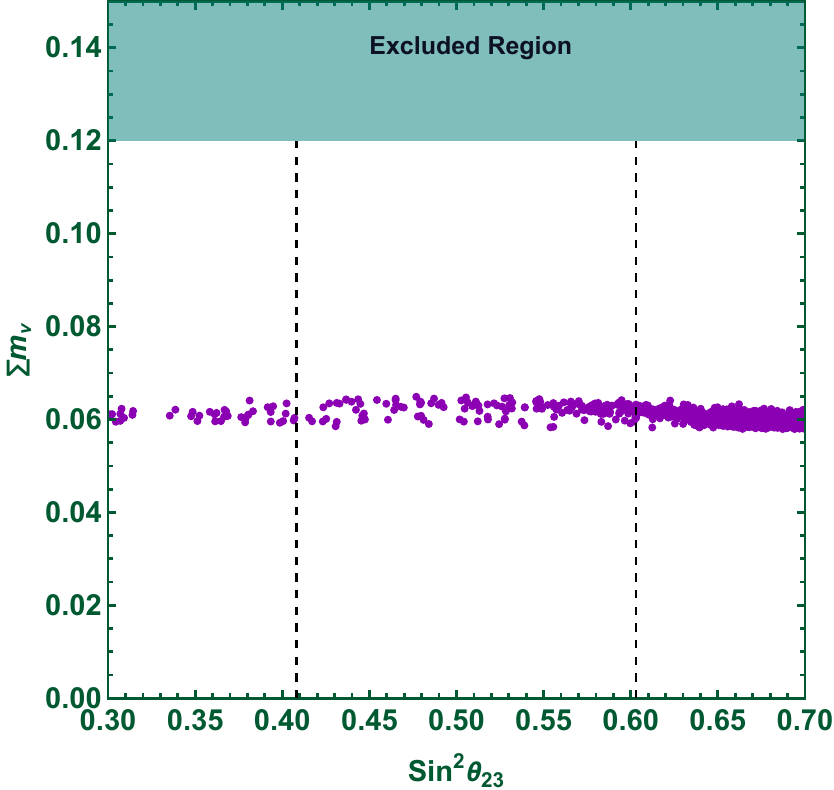}
    \caption{Correlation of the mixing angles $\sin^{2}{\theta_{13}}$, $\sin^{2}{\theta_{12}}$ and $\sin^{2}{\theta_{23}}$ with the sum of neutrino mass $\sum m_{\nu}$.}
    \label{fig:mixing angle}
\end{figure}

Fig.\ref{fig:mixing angle} illustrates the relation between the sum of neutrino masses $\sum m_{\nu}$ and the neutrino mixing angles \{$\sin^{2}\theta_{13}$,$\sin^{2}\theta_{12}$,$\sin^{2}\theta_{23}$\}. The sum of the neutrino masses is found to be below the upper bound $\sum m_{\nu} < 0.12$\, eV provided by Planck 2018\cite{Planck:2018vyg}.

\begin{figure}[!ht]
    \centering
        \centering
        \includegraphics[scale=0.47]{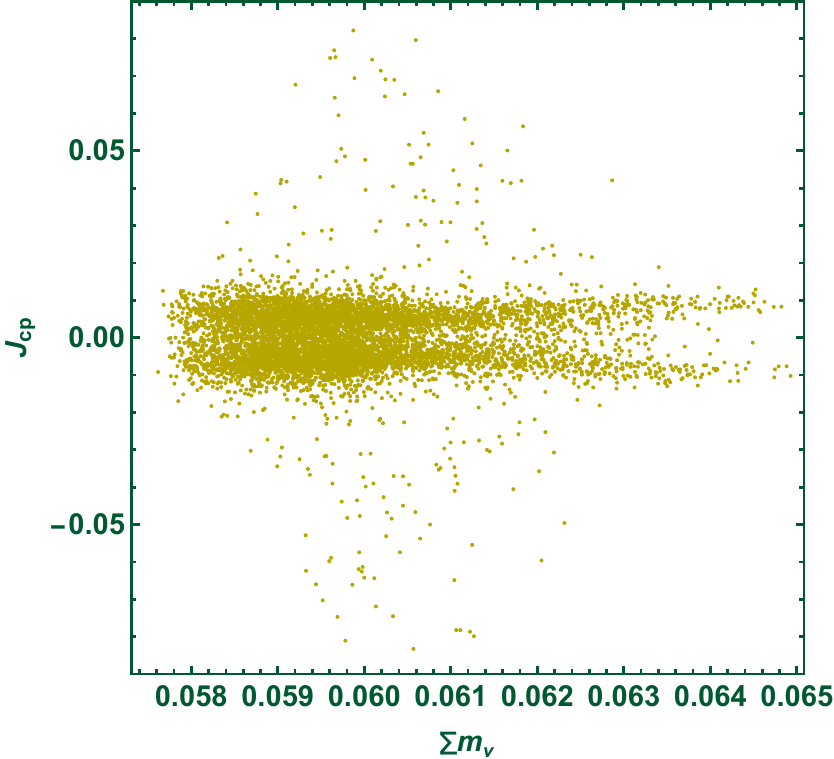}
        \hspace{2em}
        \includegraphics[scale=0.45]{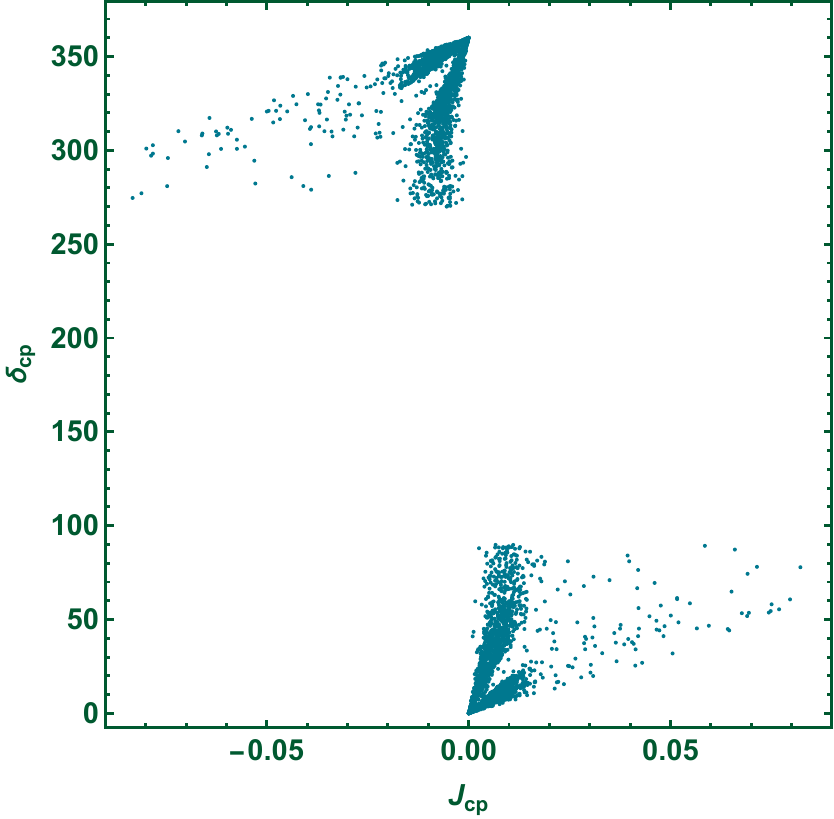}
        \caption{$(Left-panel)$ Correlation between $J_{CP}$ and sum of neutrino masses $(\sum m_{\nu})$. $(Right-panel)$ Variation of $J_{CP}$ with the Dirac CP phase $\delta_{CP}$.}
        \label{fig:fig-jcp}
\end{figure}

In Fig.\ref{fig:fig-jcp}, we have shown the correlation of Jarlskog invariant $J_{CP}$ with the sum of neutrino mass ($Left-panel$) and with the Dirac CP phase$\delta_{CP}$ ($Right-panel$). The value of $J_{CP}$ is found to be in the range $-0.08\lesssim J_{CP}\lesssim 0.08$. The value of Delta CP phase $\delta_{CP}$ is found in the range $\delta_{CP}^{o}\in[0^{\circ},89^{\circ}]$ and $\delta_{CP}^{o}\in[270^{\circ},359^{\circ}]$.



\subsection{Neutrinoless Double Beta Decay ($0 \nu \beta \beta$)}
\begin{figure}[!ht]
    \centering
        \centering
        \includegraphics[scale=0.5]{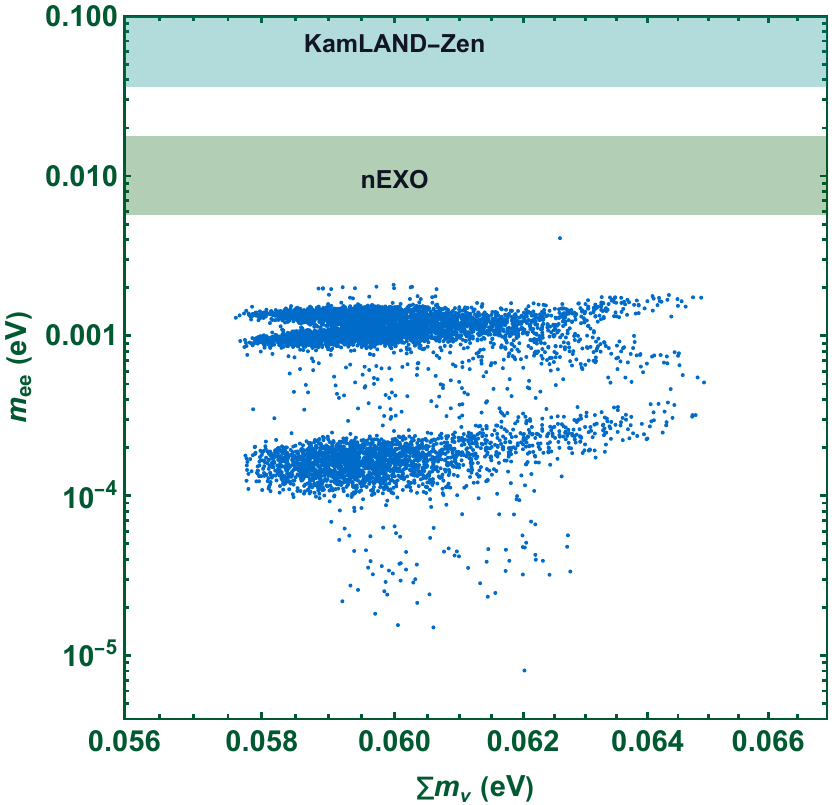}
        \caption{Effective neutrino mass $|m_{ee}|$ as a function of sum of neutrino mass ($\sum m_{\nu}$). The shaded region represents the current experimental limit on effective neutrino mass.}
        \label{fig:fig-NDBD}
\end{figure}

Neutrinoless double beta decay is a hypothetical nuclear reaction that violates conservation of lepton number by $\Delta L = 2$:
\begin{equation}
    (Z, A) \longrightarrow (Z+2,A) + 2e^{-}
\end{equation}
In this reaction, the nucleus undergoes double beta decay without emission of neutrinos. Its detection would confirm that neutrinos are Majorana particles. The decay rate for neutrinoless double beta decay is proportional to the effective neutrino mass $|m_{ee}|$:
\begin{equation}
    |m_{ee}|=|m_{1}\cos^{2}{\theta_{12}}\cos^{2}{\theta_{13}} + m_{2}\sin^{2}{\theta_{12}}\cos^{2}{\theta_{13}}e^{i \alpha_{21}} + m_{3}\sin^{2}{\theta_{13}}e^{i(\alpha_{31}-2\delta_{CP})}|
   \label{NDBD} 
\end{equation}

In Fig.\ref{fig:fig-NDBD}, we have presented our model prediction for $|m_{ee}|$ as a function of the sum of the neutrino mass $\sum m_{\nu}$. The current experimental limit from KamLAND-Zen \cite{KamLAND-Zen:2022tow} experiment for $|m_{ee}|$ is in the range $(36-156)$ meV. The sensitivity limits on $|m_{ee}|$ projected for future experiment nEXO \cite{nEXO:2017nam} lie within $(5.7-17.7)$ meV at $90\%$ C.L. Our prediction can not be explored in the current experiments, as evident from Fig.\ref {fig:fig-NDBD}.

\subsection{Baryogenesis via thermal leptogenesis}
\begin{figure}[!ht]
        \centering
        \includegraphics[scale=0.45]{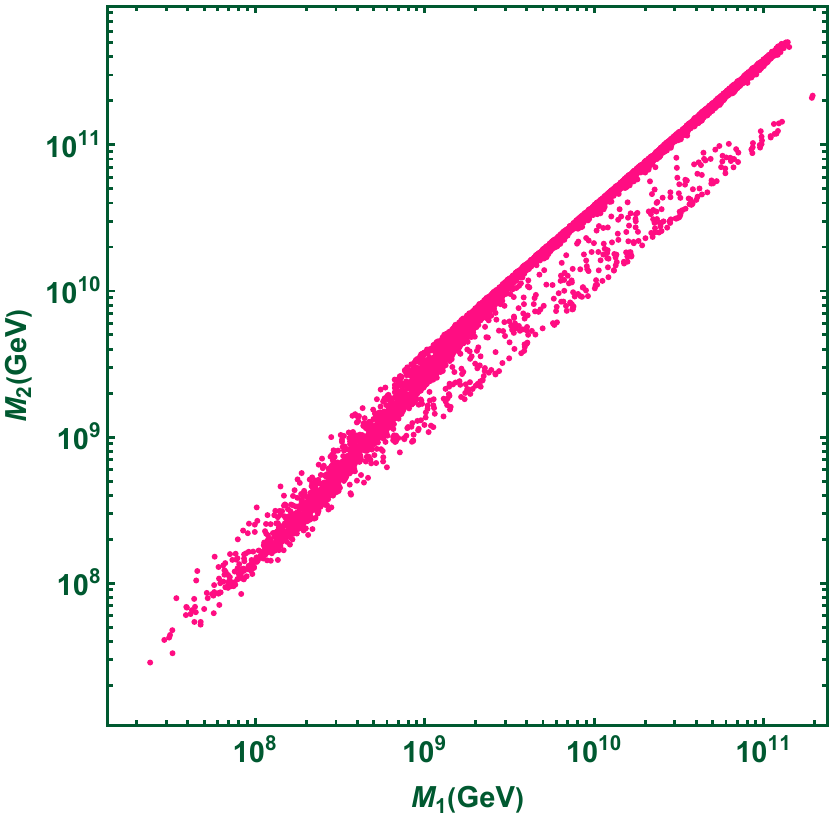}
        \hspace{1em}
        \includegraphics[scale=0.45]{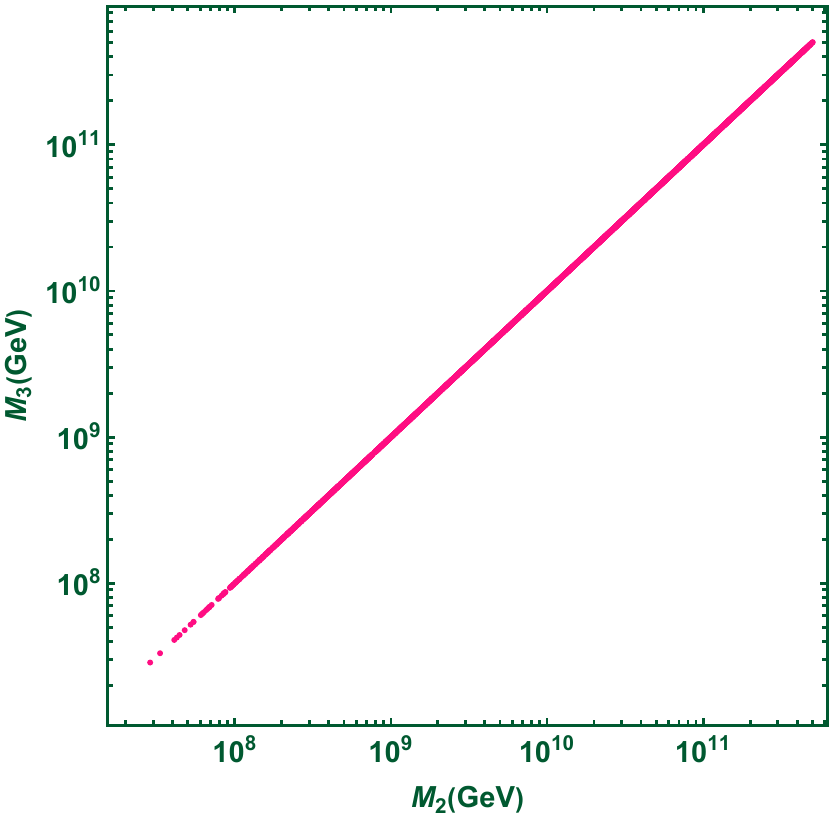}
        \caption{$(Left-panel)$ Correlation between heavy neutrino mass $M_1$ and $M_{2}$ and $(Right-panel)$ between $M_2$ and $M_3$. }
       \label{fig:Heavy neutrino mass} 
\end{figure}
In this study, we have obtained a hierarchical mass pattern for heavy right-handed neutrinos: $M_{1} < M_{2} < M_{3}$. The relation between heavy neutrino masses is illustrated in Fig.\ref{fig:Heavy neutrino mass}. The heavy neutrinos are not stable and they decay into SM lepton and Higgs doublet via Yukawa couplings\cite{Pathak:2025zdp,Marciano:2024nwm}. The decay of right-handed neutrinos violates lepton number as they can decay into both leptons and anti-leptons due to their Majorana nature. The out of equillibrium decay of right handed neutrinos produces the lepton asymmetry, which can be transferred into baryon asymmetry via the SM \textit{sphaleron process}. This is known as the leptogenesis mechanism \cite{Davidson:2008bu,Buchmuller:2004nz,Buchmuller:2005eh,Pilaftsis:1997jf,Blanchet:2010kw,Plumacher:1996kc,Barbieri:1999ma,FileviezPerez:2021hbc} for the production of matter-antimatter asymmetry in our universe. In our work, we have used the \textbf{ULYSSES}\cite{Granelli:2020pim,Granelli:2023vcm} software package to study thermal leptogenesis\footnote{For degenerate mass splitting with $|M_{i}-M_{j}|\sim \Gamma_{i}$, one has to consider the resonant leptogenesis scenario\cite{Asaka:2018hyk,Iso:2010mv, Granelli:2020ysj,Pathak:2025zdp}. Here $\Gamma_{i}$ is the decay rate of right handed neutrinos.}\cite{Das:2019ntw,Chakraborty:2019zas,Buchmuller:2004tu}. The semi-classical Boltzmann equation for one decaying right handed neutrino $N_{1}$ with number density $N_{N_{1}}$ in the single flavour approximation is given by\cite{Blanchet:2011xq}:
\begin{equation}
    \begin{split}
        \frac{dN_{N_{1}}}{dz} &= -D_{1}\bigg(N_{N_1} - N_{N_{1}}^{eq}\bigg)\\
        \frac{dN_{B-L}}{dz} &= \epsilon^{(1)} D_{1}\bigg(N_{N_{1}}-N_{N_{1}}^{eq}\bigg) - W_{1}N_{B-L}
    \end{split}
    \label{eq: BE}
\end{equation}
where $N_{B-L}$, $D_{1}$, and $W_{1}$ are the lepton asymmetry number density, decay, and washout, respectively. The evolution parameter $z=\frac{M_{1}}{T}$ where $M_{1}$ is the mass of lightes right handed neutrino and $T$ is the plasma temperature. The 
\par

The generated lepton asymmetry $N_{B-L}$ is converted into the baryon to photon ratio via the following relation:
\begin{equation}
    \eta_{B} = a_{sph}\frac{N_{B-L}}{N_{\gamma}^{rec}}=\frac{28}{79}\frac{1}{27} N_{B-L}=0.013 N_{B-L}
\end{equation}
where $a_{sph} = \frac{8}{23}$, as in our case, we assumed that both Higgs doublets survive at sphaleron freeze-out temperature\footnote{For one Higgs doublet $a_{sph}=\frac{28}{79}$, and is called the Standard model sphaleron factor. }. The $\frac{1}{27}$ factor comes from the dilution of the baryon asymmetry by photons. This quantity can be measured using two independent probes: Big-Bang nucleosynthesis (BBN) and Microwave Background radiation (CMB) data. The measured value of $\eta_{B}$ by this two mwthod is given as follows:
\begin{equation}
    \begin{split}
        \eta_{B}(BBN) &= (5.80 - 6.60) \times 10^{-10}\\
        \eta_{B}(CMB) &= (6.02 - 6.18) \times 10^{-10}
    \end{split}
    \label{eq:BAU obs_val}
\end{equation}
at $95\%$ C.L.
\par
The CP asymmetry $\epsilon_{\alpha\alpha}$ is generated through the out of equillibrium decay of the lightest right handed neutrino $N_{1}$. In the kinetic equations off diagonal flavor oscillation is being excluded. Hence, the total lepton asymmetry $\epsilon^{(1)}$ mentioned in Eq.\eqref{eq: BE} is calculated by summing over the flavor indices and is given by the following relation\cite{Das:2019ntw,Davidson:2008bu,Chakraborty:2019zas}:
\begin{equation}
    \epsilon =\sum_{\alpha}\epsilon_{\alpha\alpha} = \frac{1}{8\pi(Y_{D}^{\dagger}Y_{D})_{11}}\sum_{j}Im\bigg\{\bigg[(Y_{D}^{\dagger}Y_{D})_{1j}\bigg]^2\bigg\}g(x_{j})
\end{equation}
where $x_{j}=\frac{M_{j}^2}{M_{1}^2}$ and 
\begin{equation}
   g(x) = \sqrt{x}\bigg[\frac{1}{1-x}+1-(1+x)ln\bigg(\frac{1+x}{x}\bigg)\bigg]
\end{equation}
The input Yukawa matrix for the calculation of baryon asymmetry in the \textbf{ULYSSES} package is given below:
\begin{equation}
Y_{D}=
    \begin{pmatrix}
      -9.4\times10^{-5}+i 2.7\times10^{-5} & -5.9\times10^{-4} + i 3\times10^{-3} & 4.02\times10^{-3} + i2.3\times10^{-3}\\
      1.5\times10^{-3}-i6.05\times10^{-3} & 2.4\times10^{-6}+i1.3\times10^{-4} & -4.2\times10^{-4} - i1.1\times10^{-3}\\
      -2.5\times10^{-3}+i2.04\times10^{-4} & 3.8\times10^{-3}+i2.2\times10^{-3} & 5.5\times10^{-5}-i1.5\times10^{-4}
    \end{pmatrix}
\end{equation}
The mass of heavy right handed neutrinos taken as $M_{1}=1.8\times10^{10}GeV, M_{2}=6.7\times10^{10}GeV, M_{3}=1.2\times10^{11}GeV$. The right-handed neutrino mass and Yukawa Couplings also satisfy the neutrino oscillation data, as discussed earlier. We have taken only one set of these values to study the baryon asymmetry using the \textbf{ULYSSES} software package.
\begin{figure}
    \centering
        \centering
        \includegraphics[scale=1]{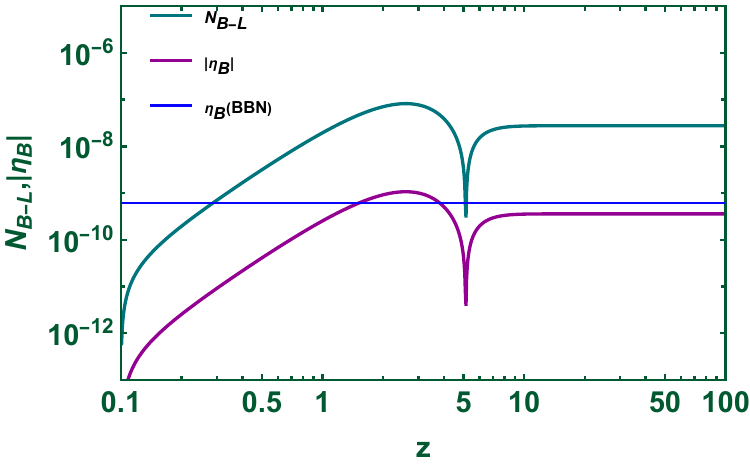}
        \caption{Variation of lepton number density for single flavor and baryon asymmetry parameter with $z=\frac{M_1}{T}$. The Cyan, Purple, and Blue lines corresponding to $N_{B-L}$, $\eta_{B}$ and $\eta_{B}(BBN)$ respectively.}
        \label{fig:fig-BAU}
\end{figure}
From Fig.\ref{fig:fig-BAU}, we can see the evolution of single flavor lepton number density \textit{(Cyan line)} and baryon asymmetry \textit{(Purple line)} with $z=\frac{M_{1}}{T}$. The observed value of $\eta_{B}$ given in Eq.\eqref{eq:BAU obs_val}  is found to be in close agreement with the value predicted by our model.

\section{Conclusion}\label{sec:5}

In this work, we have developed a model to study neutrino phenomenology using FN-like modular symmetry. We have implemented $T'$ modular symmetry, and neutrino mass is generated via the Type I seesaw framework. FN-like modular symmetry work has been done recently for the quark sector\cite{Kuranaga:2021ujd}. In this work, we have implemented the same mechanism for the neutrino sector. The advantage of using FN-like modular symmetry is that we do not need additional symmetry $U(1)_{FN}$, unlike the other FN mechanism without modular symmetry. Here, the modular weights play the role of $U(1)_{FN}$ charge. A scalar field with a negative modular weight has been introduced, and the allowed Yukawa couplings are then suppressed by powers of the scalar VEV ($\langle \phi \rangle/\Lambda$). Our model produces the mass-squared difference of neutrino mass within the allowed region. All the mixing angles are also found within the $3\sigma$ bound. The sum of the light neutrino mass is found within  $\sum m_{\nu} \in [0.0576-0.0646]$ eV in NO. We have also carried out a study of neutrinoless double beta decay, and the $m_{ee}$ value is obtained below the current experiment sensitivity limit for our model. We have also investigated the baryon asymmetry of our universe for the model. We studied baryogenesis via the thermal leptogenesis mechanism for the single flavor case in our work, and it is found to be consistent with the current observed value. It will be interesting to see if this mechanism can be realized in other neutrino mass models also. 


\section*{Acknowledgement}
GP would like to acknowledge CSIR-HRDG for the financial support received in the form of SRF fellowship (09/0796(16046)/2022-EMR-I). 



\bibliographystyle{JHEP}
\bibliography{reference}

\providecommand{\href}[2]{#2}\begingroup\raggedright\begin{thebibliography}{10}

\bibitem{Feruglio:2017spp}
F.~Feruglio, {\em {Are neutrino masses modular forms?}}, pp.~227--266.
\newblock 2019.
\newblock \href{http://arxiv.org/abs/1706.08749}{{\tt arXiv:1706.08749}}.

\bibitem{Meloni:2023aru}
D.~Meloni and M.~Parriciatu, {\it {A simplest modular S$_{3}$ model for leptons}},  {\em JHEP} {\bf 09} (2023) 043, [\href{http://arxiv.org/abs/2306.09028}{{\tt arXiv:2306.09028}}].

\bibitem{Okada:2019xqk}
H.~Okada and Y.~Orikasa, {\it {Modular $S_3$ symmetric radiative seesaw model}},  {\em Phys. Rev. D} {\bf 100} (2019), no.~11 115037, [\href{http://arxiv.org/abs/1907.04716}{{\tt arXiv:1907.04716}}].

\bibitem{Kashav:2021zir}
M.~Kashav and S.~Verma, {\it {Broken scaling neutrino mass matrix and leptogenesis based on A$_{4}$ modular invariance}},  {\em JHEP} {\bf 09} (2021) 100, [\href{http://arxiv.org/abs/2103.07207}{{\tt arXiv:2103.07207}}].

\bibitem{Nomura:2019jxj}
T.~Nomura and H.~Okada, {\it {A modular $A_4$ symmetric model of dark matter and neutrino}},  {\em Phys. Lett. B} {\bf 797} (2019) 134799, [\href{http://arxiv.org/abs/1904.03937}{{\tt arXiv:1904.03937}}].

\bibitem{Behera:2020lpd}
M.~K. Behera, S.~Singirala, S.~Mishra, and R.~Mohanta, {\it {A modular A $_{4}$ symmetric scotogenic model for neutrino mass and dark matter}},  {\em J. Phys. G} {\bf 49} (2022), no.~3 035002, [\href{http://arxiv.org/abs/2009.01806}{{\tt arXiv:2009.01806}}].

\bibitem{Altarelli:2005yx}
G.~Altarelli and F.~Feruglio, {\it {Tri-bimaximal neutrino mixing, A(4) and the modular symmetry}},  {\em Nucl. Phys. B} {\bf 741} (2006) 215--235, [\href{http://arxiv.org/abs/hep-ph/0512103}{{\tt hep-ph/0512103}}].

\bibitem{Pathak:2025zdp}
G.~Pathak and M.~K. Das, {\it {Matter-antimatter asymmetry in minimal inverse seesaw framework with $A_4$ modular symmetry}},  \href{http://arxiv.org/abs/2505.03000}{{\tt arXiv:2505.03000}}.

\bibitem{Pathak:2024sei}
G.~Pathak, P.~Das, and M.~K. Das, {\it {Neutrino mass genesis in scoto-inverse seesaw with modular $A_4$}},  {\em Eur. Phys. J. C} {\bf 85} (2025), no.~5 569, [\href{http://arxiv.org/abs/2411.13895}{{\tt arXiv:2411.13895}}].

\bibitem{Kobayashi:2019xvz}
T.~Kobayashi, Y.~Shimizu, K.~Takagi, M.~Tanimoto, and T.~H. Tatsuishi, {\it {$A_4$ lepton flavor model and modulus stabilization from $S_4$ modular symmetry}},  {\em Phys. Rev. D} {\bf 100} (2019), no.~11 115045, [\href{http://arxiv.org/abs/1909.05139}{{\tt arXiv:1909.05139}}]. [Erratum: Phys.Rev.D 101, 039904 (2020)].

\bibitem{Penedo:2018nmg}
J.~T. Penedo and S.~T. Petcov, {\it {Lepton Masses and Mixing from Modular $S_4$ Symmetry}},  {\em Nucl. Phys. B} {\bf 939} (2019) 292--307, [\href{http://arxiv.org/abs/1806.11040}{{\tt arXiv:1806.11040}}].

\bibitem{Zhang:2021olk}
X.~Zhang and S.~Zhou, {\it {Inverse seesaw model with a modular S 4 symmetry: lepton flavor mixing and warm dark~matter}},  {\em JCAP} {\bf 09} (2021) 043, [\href{http://arxiv.org/abs/2106.03433}{{\tt arXiv:2106.03433}}].

\bibitem{Wang:2019ovr}
X.~Wang and S.~Zhou, {\it {The minimal seesaw model with a modular S$_{4}$ symmetry}},  {\em JHEP} {\bf 05} (2020) 017, [\href{http://arxiv.org/abs/1910.09473}{{\tt arXiv:1910.09473}}].

\bibitem{Novichkov:2018nkm}
P.~P. Novichkov, J.~T. Penedo, S.~T. Petcov, and A.~V. Titov, {\it {Modular A$_{5}$ symmetry for flavour model building}},  {\em JHEP} {\bf 04} (2019) 174, [\href{http://arxiv.org/abs/1812.02158}{{\tt arXiv:1812.02158}}].

\bibitem{Ding:2019xna}
G.-J. Ding, S.~F. King, and X.-G. Liu, {\it {Neutrino mass and mixing with $A_5$ modular symmetry}},  {\em Phys. Rev. D} {\bf 100} (2019), no.~11 115005, [\href{http://arxiv.org/abs/1903.12588}{{\tt arXiv:1903.12588}}].

\bibitem{Liu:2019khw}
X.-G. Liu and G.-J. Ding, {\it {Neutrino Masses and Mixing from Double Covering of Finite Modular Groups}},  {\em JHEP} {\bf 08} (2019) 134, [\href{http://arxiv.org/abs/1907.01488}{{\tt arXiv:1907.01488}}].

\bibitem{Mishra:2023cjc}
P.~Mishra, M.~K. Behera, and R.~Mohanta, {\it {Neutrino phenomenology, W-mass anomaly, and muon (g-2) in a minimal type-III seesaw model using a T' modular symmetry}},  {\em Phys. Rev. D} {\bf 107} (2023), no.~11 115004, [\href{http://arxiv.org/abs/2302.00494}{{\tt arXiv:2302.00494}}].

\bibitem{Novichkov:2020eep}
P.~P. Novichkov, J.~T. Penedo, and S.~T. Petcov, {\it {Double cover of modular $S_4$ for flavour model building}},  {\em Nucl. Phys. B} {\bf 963} (2021) 115301, [\href{http://arxiv.org/abs/2006.03058}{{\tt arXiv:2006.03058}}].

\bibitem{Liu:2020akv}
X.-G. Liu, C.-Y. Yao, and G.-J. Ding, {\it {Modular invariant quark and lepton models in double covering of $S_4$ modular group}},  {\em Phys. Rev. D} {\bf 103} (2021), no.~5 056013, [\href{http://arxiv.org/abs/2006.10722}{{\tt arXiv:2006.10722}}].

\bibitem{Wang:2020lxk}
X.~Wang, B.~Yu, and S.~Zhou, {\it {Double covering of the modular $A_5$ group and lepton flavor mixing in the minimal seesaw model}},  {\em Phys. Rev. D} {\bf 103} (2021), no.~7 076005, [\href{http://arxiv.org/abs/2010.10159}{{\tt arXiv:2010.10159}}].

\bibitem{Yao:2020zml}
C.-Y. Yao, X.-G. Liu, and G.-J. Ding, {\it {Fermion masses and mixing from the double cover and metaplectic cover of the $A_5$ modular group}},  {\em Phys. Rev. D} {\bf 103} (2021), no.~9 095013, [\href{http://arxiv.org/abs/2011.03501}{{\tt arXiv:2011.03501}}].

\bibitem{Behera:2021eut}
M.~K. Behera and R.~Mohanta, {\it {Inverse seesaw in $A_5^\prime$ modular symmetry}},  {\em J. Phys. G} {\bf 49} (2022), no.~4 045001, [\href{http://arxiv.org/abs/2108.01059}{{\tt arXiv:2108.01059}}].

\bibitem{Froggatt:1978nt}
C.~D. Froggatt and H.~B. Nielsen, {\it {Hierarchy of Quark Masses, Cabibbo Angles and CP Violation}},  {\em Nucl. Phys. B} {\bf 147} (1979) 277--298.

\bibitem{Ibe:2024cvi}
M.~Ibe, S.~Shirai, and K.~Watanabe, {\it {Comprehensive Bayesian exploration of Froggatt-Nielsen mechanism}},  {\em JHEP} {\bf 03} (2025) 150, [\href{http://arxiv.org/abs/2412.19484}{{\tt arXiv:2412.19484}}].

\bibitem{Kamikado:2008jx}
H.~Kamikado, T.~Shindou, and E.~Takasugi, {\it {Froggatt-Nielsen hierarchy and the neutrino mass matrix}},  \href{http://arxiv.org/abs/0805.1338}{{\tt arXiv:0805.1338}}.

\bibitem{Qiu:2023igq}
Y.-C. Qiu, J.-W. Wang, and T.~T. Yanagida, {\it {Predictions of mee and neutrino mass from a consistent Froggatt-Nielsen model}},  {\em Phys. Rev. D} {\bf 108} (2023), no.~11 115021, [\href{http://arxiv.org/abs/2307.16470}{{\tt arXiv:2307.16470}}].

\bibitem{Cornella:2024jaw}
C.~Cornella, D.~Curtin, G.~Krnjaic, and M.~Mellors, {\it {Testing the Froggatt-Nielsen Mechanism with Lepton Violation}},  \href{http://arxiv.org/abs/2501.00629}{{\tt arXiv:2501.00629}}.

\bibitem{Brdar:2019iem}
V.~Brdar, A.~J. Helmboldt, S.~Iwamoto, and K.~Schmitz, {\it {Type-I Seesaw as the Common Origin of Neutrino Mass, Baryon Asymmetry, and the Electroweak Scale}},  {\em Phys. Rev. D} {\bf 100} (2019) 075029, [\href{http://arxiv.org/abs/1905.12634}{{\tt arXiv:1905.12634}}].

\bibitem{Behera:2024vfv}
M.~K. Behera, P.~Ittisamai, C.~Pongkitivanichkul, and P.~Uttayarat, {\it {Exploring type-I seesaw under S 3 modular symmetry}},  {\em EPJ Web Conf.} {\bf 312} (2024) 02010.

\bibitem{minkowski1977mu}
P.~Minkowski, {\it $\mu$→ e$\gamma$ at a rate of one out of 109 muon decays?},  {\em Physics Letters B} {\bf 67} (1977), no.~4 421--428.

\bibitem{schechter1980neutrino}
J.~Schechter and J.~W. Valle, {\it Neutrino masses in $su (2) \otimes u (1)$ theories},  {\em Physical Review D} {\bf 22} (1980), no.~9 2227.

\bibitem{mohapatra1980neutrino}
R.~N. Mohapatra and G.~Senjanovi{\'c}, {\it Neutrino mass and spontaneous parity nonconservation},  {\em Physical Review Letters} {\bf 44} (1980), no.~14 912.

\bibitem{King:2020qaj}
S.~J.~D. King and S.~F. King, {\it {Fermion mass hierarchies from modular symmetry}},  {\em JHEP} {\bf 09} (2020) 043, [\href{http://arxiv.org/abs/2002.00969}{{\tt arXiv:2002.00969}}].

\bibitem{Kuranaga:2021ujd}
H.~Kuranaga, H.~Ohki, and S.~Uemura, {\it {Modular origin of mass hierarchy: Froggatt-Nielsen like mechanism}},  {\em JHEP} {\bf 07} (2021) 068, [\href{http://arxiv.org/abs/2105.06237}{{\tt arXiv:2105.06237}}].

\bibitem{Jones:2021cga}
B.~J.~P. Jones, {\it {The Physics of Neutrinoless Double Beta Decay: A Primer}},  in {\em {Theoretical Advanced Study Institute in Elementary Particle Physics}: {The Obscure Universe: Neutrinos and Other Dark Matters}}, 8, 2021.
\newblock \href{http://arxiv.org/abs/2108.09364}{{\tt arXiv:2108.09364}}.

\bibitem{Dolinski:2019nrj}
M.~J. Dolinski, A.~W.~P. Poon, and W.~Rodejohann, {\it {Neutrinoless Double-Beta Decay: Status and Prospects}},  {\em Ann. Rev. Nucl. Part. Sci.} {\bf 69} (2019) 219--251, [\href{http://arxiv.org/abs/1902.04097}{{\tt arXiv:1902.04097}}].

\bibitem{Bilenky:2012qi}
S.~M. Bilenky and C.~Giunti, {\it {Neutrinoless double-beta decay: A brief review}},  {\em Mod. Phys. Lett. A} {\bf 27} (2012) 1230015, [\href{http://arxiv.org/abs/1203.5250}{{\tt arXiv:1203.5250}}].

\bibitem{Gomez-Cadenas:2010zcc}
J.~J. Gomez-Cadenas, J.~Martin-Albo, M.~Sorel, P.~Ferrario, F.~Monrabal, J.~Munoz-Vidal, P.~Novella, and A.~Poves, {\it {Sense and sensitivity of double beta decay experiments}},  {\em JCAP} {\bf 06} (2011) 007, [\href{http://arxiv.org/abs/1010.5112}{{\tt arXiv:1010.5112}}].

\bibitem{Sakharov:1967dj}
A.~D. Sakharov, {\it {Violation of CP Invariance, C asymmetry, and baryon asymmetry of the universe}},  {\em Pisma Zh. Eksp. Teor. Fiz.} {\bf 5} (1967) 32--35.

\bibitem{Ding:2022aoe}
G.-J. Ding, F.~R. Joaquim, and J.-N. Lu, {\it {Texture-zero patterns of lepton mass matrices from modular symmetry}},  {\em JHEP} {\bf 03} (2023) 141, [\href{http://arxiv.org/abs/2211.08136}{{\tt arXiv:2211.08136}}].

\bibitem{hochmuth2007upmns}
K.~Hochmuth, S.~Petcov, and W.~Rodejohann, {\it $u_{PMNS}= u_{l}^{\dagger} u_{\nu}$},  {\em Physics Letters B} {\bf 654} (2007), no.~5-6 177--188.

\bibitem{Esteban:2020cvm}
I.~Esteban, M.~C. Gonzalez-Garcia, M.~Maltoni, T.~Schwetz, and A.~Zhou, {\it {The fate of hints: updated global analysis of three-flavor neutrino oscillations}},  {\em JHEP} {\bf 09} (2020) 178, [\href{http://arxiv.org/abs/2007.14792}{{\tt arXiv:2007.14792}}].

\bibitem{Antusch:2013jca}
S.~Antusch and V.~Maurer, {\it {Running quark and lepton parameters at various scales}},  {\em JHEP} {\bf 11} (2013) 115, [\href{http://arxiv.org/abs/1306.6879}{{\tt arXiv:1306.6879}}].

\bibitem{Planck:2018vyg}
{\bf Planck} Collaboration, N.~Aghanim et~al., {\it {Planck 2018 results. VI. Cosmological parameters}},  {\em Astron. Astrophys.} {\bf 641} (2020) A6, [\href{http://arxiv.org/abs/1807.06209}{{\tt arXiv:1807.06209}}]. [Erratum: Astron.Astrophys. 652, C4 (2021)].

\bibitem{KamLAND-Zen:2022tow}
{\bf KamLAND-Zen} Collaboration, S.~Abe et~al., {\it {Search for the Majorana Nature of Neutrinos in the Inverted Mass Ordering Region with KamLAND-Zen}},  {\em Phys. Rev. Lett.} {\bf 130} (2023), no.~5 051801, [\href{http://arxiv.org/abs/2203.02139}{{\tt arXiv:2203.02139}}].

\bibitem{nEXO:2017nam}
{\bf nEXO} Collaboration, J.~B. Albert et~al., {\it {Sensitivity and Discovery Potential of nEXO to Neutrinoless Double Beta Decay}},  {\em Phys. Rev. C} {\bf 97} (2018), no.~6 065503, [\href{http://arxiv.org/abs/1710.05075}{{\tt arXiv:1710.05075}}].

\bibitem{Marciano:2024nwm}
S.~Marciano, D.~Meloni, and M.~Parriciatu, {\it {Minimal seesaw and leptogenesis with the smallest modular finite group}},  {\em JHEP} {\bf 05} (2024) 020, [\href{http://arxiv.org/abs/2402.18547}{{\tt arXiv:2402.18547}}].

\bibitem{Davidson:2008bu}
S.~Davidson, E.~Nardi, and Y.~Nir, {\it {Leptogenesis}},  {\em Phys. Rept.} {\bf 466} (2008) 105--177, [\href{http://arxiv.org/abs/0802.2962}{{\tt arXiv:0802.2962}}].

\bibitem{Buchmuller:2004nz}
W.~Buchmuller, P.~Di~Bari, and M.~Plumacher, {\it {Leptogenesis for pedestrians}},  {\em Annals Phys.} {\bf 315} (2005) 305--351, [\href{http://arxiv.org/abs/hep-ph/0401240}{{\tt hep-ph/0401240}}].

\bibitem{Buchmuller:2005eh}
W.~Buchmuller, R.~D. Peccei, and T.~Yanagida, {\it {Leptogenesis as the origin of matter}},  {\em Ann. Rev. Nucl. Part. Sci.} {\bf 55} (2005) 311--355, [\href{http://arxiv.org/abs/hep-ph/0502169}{{\tt hep-ph/0502169}}].

\bibitem{Pilaftsis:1997jf}
A.~Pilaftsis, {\it {CP violation and baryogenesis due to heavy Majorana neutrinos}},  {\em Phys. Rev. D} {\bf 56} (1997) 5431--5451, [\href{http://arxiv.org/abs/hep-ph/9707235}{{\tt hep-ph/9707235}}].

\bibitem{Blanchet:2010kw}
S.~Blanchet, P.~S.~B. Dev, and R.~N. Mohapatra, {\it {Leptogenesis with TeV Scale Inverse Seesaw in SO(10)}},  {\em Phys. Rev. D} {\bf 82} (2010) 115025, [\href{http://arxiv.org/abs/1010.1471}{{\tt arXiv:1010.1471}}].

\bibitem{Plumacher:1996kc}
M.~Plumacher, {\it {Baryogenesis and lepton number violation}},  {\em Z. Phys. C} {\bf 74} (1997) 549--559, [\href{http://arxiv.org/abs/hep-ph/9604229}{{\tt hep-ph/9604229}}].

\bibitem{Barbieri:1999ma}
R.~Barbieri, P.~Creminelli, A.~Strumia, and N.~Tetradis, {\it {Baryogenesis through leptogenesis}},  {\em Nucl. Phys. B} {\bf 575} (2000) 61--77, [\href{http://arxiv.org/abs/hep-ph/9911315}{{\tt hep-ph/9911315}}].

\bibitem{FileviezPerez:2021hbc}
P.~Fileviez~Perez, C.~Murgui, and A.~D. Plascencia, {\it {Baryogenesis via leptogenesis: Spontaneous B and L violation}},  {\em Phys. Rev. D} {\bf 104} (2021), no.~5 055007, [\href{http://arxiv.org/abs/2103.13397}{{\tt arXiv:2103.13397}}].

\bibitem{Granelli:2020pim}
A.~Granelli, K.~Moffat, Y.~F. Perez-Gonzalez, H.~Schulz, and J.~Turner, {\it {ULYSSES: Universal LeptogeneSiS Equation Solver}},  {\em Comput. Phys. Commun.} {\bf 262} (2021) 107813, [\href{http://arxiv.org/abs/2007.09150}{{\tt arXiv:2007.09150}}].

\bibitem{Granelli:2023vcm}
A.~Granelli, C.~Leslie, Y.~F. Perez-Gonzalez, H.~Schulz, B.~Shuve, J.~Turner, and R.~Walker, {\it {ULYSSES, universal LeptogeneSiS equation solver: Version 2}},  {\em Comput. Phys. Commun.} {\bf 291} (2023) 108834, [\href{http://arxiv.org/abs/2301.05722}{{\tt arXiv:2301.05722}}].

\bibitem{Asaka:2018hyk}
T.~Asaka and T.~Yoshida, {\it {Resonant leptogenesis at TeV-scale and neutrinoless double beta decay}},  {\em JHEP} {\bf 09} (2019) 089, [\href{http://arxiv.org/abs/1812.11323}{{\tt arXiv:1812.11323}}].

\bibitem{Iso:2010mv}
S.~Iso, N.~Okada, and Y.~Orikasa, {\it {Resonant Leptogenesis in the Minimal B-L Extended Standard Model at TeV}},  {\em Phys. Rev. D} {\bf 83} (2011) 093011, [\href{http://arxiv.org/abs/1011.4769}{{\tt arXiv:1011.4769}}].

\bibitem{Granelli:2020ysj}
A.~Granelli, K.~Moffat, and S.~T. Petcov, {\it {Flavoured resonant leptogenesis at sub-TeV scales}},  {\em Nucl. Phys. B} {\bf 973} (2021) 115597, [\href{http://arxiv.org/abs/2009.03166}{{\tt arXiv:2009.03166}}].

\bibitem{Das:2019ntw}
P.~Das, M.~K. Das, and N.~Khan, {\it {Phenomenological study of neutrino mass, dark matter and baryogenesis within the framework of minimal extended seesaw}},  {\em JHEP} {\bf 03} (2020) 018, [\href{http://arxiv.org/abs/1911.07243}{{\tt arXiv:1911.07243}}].

\bibitem{Chakraborty:2019zas}
I.~Chakraborty and H.~Roy, {\it {Type-I thermal leptogenesis in $Z_3$-symmetric three Higgs doublet model}},  {\em Eur. Phys. J. C} {\bf 80} (2020), no.~11 1038, [\href{http://arxiv.org/abs/1909.07790}{{\tt arXiv:1909.07790}}].

\bibitem{Buchmuller:2004tu}
W.~Buchmuller, P.~Di~Bari, and M.~Plumacher, {\it {Some aspects of thermal leptogenesis}},  {\em New J. Phys.} {\bf 6} (2004) 105, [\href{http://arxiv.org/abs/hep-ph/0406014}{{\tt hep-ph/0406014}}].

\bibitem{Blanchet:2011xq}
S.~Blanchet, P.~Di~Bari, D.~A. Jones, and L.~Marzola, {\it {Leptogenesis with heavy neutrino flavours: from density matrix to Boltzmann equations}},  {\em JCAP} {\bf 01} (2013) 041, [\href{http://arxiv.org/abs/1112.4528}{{\tt arXiv:1112.4528}}].

\end{thebibliography}\endgroup
\end{document}